\documentclass[fleqn,twoside]{article}
\usepackage{espcrc2}

\newcommand{\<}{\langle}
\renewcommand{\>}{\rangle} 
\newcommand{\beq}{\begin{equation}}
\newcommand{\eeq}{\end{equation}}
\newcommand{\beqn}{\begin{eqnarray}}
\newcommand{\eeqn}{\end{eqnarray}}

\newcommand{\AmS}{{\protect\the\textfont2
A\kern-.1667em\lower.5ex\hbox{M}\kern-.125emS}}

\title{Twisted-mass lattice QCD with mass non-degenerate quarks
\thanks{Talk presented by G.C. Rossi.}}

\author{R. Frezzotti\address{INFN, Sezione di Milano and 
Dipartimento di Fisica, Universit\`a di Milano ``{\it Bicocca}''\\
Piazza della Scienza 3 - 20126 Milano (Italy)}
and G.C. Rossi\address{Dipartimento 
di Fisica, Universit\`a di  Roma ``{\it Tor Vergata}'' and INFN, 
Sezione di Roma 2\\Via della Ricerca Scientifica - 00133 Roma (Italy)}}

\begin{document}

\begin{abstract}
The maximally twisted lattice QCD action of an $SU_f(2)$ doublet 
of mass degenerate Wilson quarks gives rise to a real positive fermion 
determinant and it is invariant under the product of standard parity times 
the change of sign of the coefficient of the Wilson term. The existence 
of this spurionic symmetry implies that O($a$) improvement is either automatic
or achieved through simple linear combinations of quantities taken with 
opposite external three-momenta. We show that in the case of maximal twist 
all these nice results can be extended to the more interesting case of a mass 
non-degenerate quark pair.
\vspace{1pc}
\end{abstract}
\maketitle

\section{Introduction} \label{sec:INTRO}

It has been shown in ref.~\cite{FR} that in lattice QCD (LQCD) with mass 
degenerate $SU_f(2)$ doublets of Wilson fermions it is possible to improve the 
approach to the continuum limit of correlation functions of gauge invariant 
multiplicative renormalizable (m.r.) operators by taking the arithmetic 
average (Wilson average - $WA$) of pairs of correlators computed in theories 
regularized with Wilson terms of opposite sign and identical values of 
the subtracted (unrenormalized) lattice quark mass, $m^W_q=M_0-M_{\rm{cr}}$ 
with $M_0$ the bare quark mass. Equivalently one can take appropriate linear 
combinations of the correlators computed in a given regularization (fixed 
sign of the Wilson term) but with opposite values of $m^W_q$ (mass average - 
$MA$). Improved hadronic masses and matrix elements can be similarly 
obtained by taking $WA$'s of the corresponding quantities separately 
computed within the two regularizations.

To avoid the difficulties related to the nature of the spectrum of the 
Wilson--Dirac (WD) operator (here we are referring to the well-known problem 
of the existence of ``exceptional configurations''~\cite{EXCONF,IRV}) 
twisted-mass lattice QCD (tm-LQCD)~\cite{TM} should be used for the actual 
computation of the correlators taking part to the averages. The fermionic 
determinant of tm-LQCD is, in fact, real and positive on arbitrary gauge 
backgrounds, as long as the quark mass is non-vanishing. All the nice 
cancellations of O($a$) terms that one finds in the standard Wilson case 
extend to tm-LQCD with mass degenerate quark doublets. 

Peculiar simplifications occur if the twisting angle is taken to be equal 
to $\pi/2$ (maximal twist). This choice is particularly useful for 
applications because essentially all interesting physical quantities (e.g.\ 
hadronic masses and matrix elements of m.r.\ operators between states with 
vanishing three-momenta) can be extracted from O($a$) improved lattice data 
without making recourse to any $WA$.

The main goal of this note is to show that in the maximally twisted case 
positivity of the determinant and O($a$) improvement without $WA$ can
be extended to encompass the more interesting situation in which mass 
non-degenerate quark pairs are considered. This is a preliminary, necessary 
step if one wants to set up a realistic computational scheme for operator 
matrix elements.

For future applications it is important to stress that the approach we 
propose is very flexible as it allows to regularize different flavours
with differently twisted Wilson terms. As discussed in detail in~\cite{FR2}, 
it is possible to exploit this freedom in order to improve the chiral 
behaviour of lattice correlators to the point of (hopefully) killing all the 
unwanted ``wrong chirality mixings'' that affect in the standard 
Wilson~\cite{BMMRT} and twisted-mass~\cite{TM,GHPSV} regularizations of 
QCD the construction of m.r.\ operators. Of special importance is the 
application of this new strategy to the construction of the m.r.\ operators 
which represent the CP-conserving, $\Delta S=2$ and $\Delta S=1$ effective
weak Hamiltonian on the lattice.

The presentation of the material is divided in two parts. In the first 
part (sect.~\ref{sec:SUMMARY}) we spell out in detail the formulae 
relevant for mass degenerate quark pairs with the choice $\pm\pi/2$ (maximal 
twist) for the twisting angle. In the second part (sect.~\ref{sec:MNDQ}) 
we discuss how one can deal with mass non-degenerate pairs of quarks without 
loosing O($a$) improvement, or the positivity of the fermion determinant. The 
latter property, which is obvious in the case of mass degenerate doublets, 
is crucial for being able to actually carry out numerical Monte Carlo 
simulations. Conclusions can be found in sect.~\ref{sec:CONCL}. In Appendix~A 
we prove the renormalizability of the maximally twisted fermion action both 
in the case of mass degenerate and mass non-degenerate quarks. The proof of 
the positivity of the fermion determinant in the non-trivial case of mass 
non-degenerate quarks is given in Appendix~B.

\section{A summary of tm-LQCD @ $\pm \pi/2$} \label{sec:SUMMARY}

The twisted-mass lattice action of an $SU_f(2)$ flavour doublet of 
mass degenerate quark has the form
\beqn
&&S_{\rm{F,D}}^{(\omega)}[\psi,\bar\psi,U]=
a^4 \sum_x\,\bar\psi(x) 
\Big{[}\gamma\cdot\widetilde\nabla+
\label{PHYSCHI}\\&&+e^{-i\omega\gamma_5\tau_3}
W_{\rm{cr}}(r)+m_q\Big{]}\psi(x)\, ,\nonumber\eeqn
where 
\beqn 
&&\gamma\cdot\widetilde\nabla\equiv
\frac{1}{2}\sum_\mu\gamma_\mu(\nabla^\star_\mu+\nabla_\mu)\, ,\label{WDONDD}\\
&&W_{\rm{cr}}(r)\equiv-a\frac{r}{2}\sum_\mu\nabla^\star_\mu\nabla_\mu+
M_{\rm{cr}}(r)\label{WDONDM}\eeqn
and $M_{\rm{cr}}(r)=-M_{\rm{cr}}(-r)$ is the critical quark mass. We wrote 
the action in what is usually called the ``physical basis''~\cite{FR}, where 
$m_q$, is taken to be real (and positive)~\footnote{Unless differently 
stated, we employ here the notations of ref.~\cite{FR}.}. By undoing the 
twisting of the Wilson term and bringing it fully to the mass term, it was 
proved in ref.~\cite{FR} that the parameter $M_{\rm{cr}}(r)$ appearing in the 
action~(\ref{PHYSCHI}) is equal to that of the corresponding standard Wilson 
theory.

In this note we will concentrate on the maximal twist case 
$\omega=\pi/2$, where eq.~(\ref{PHYSCHI}) becomes~\footnote{Whatever
we say for $\omega=\pi/2$ also holds for $\omega=-\pi/2$.} 
\beqn
&&S_{\rm{F,D}}^{(\pi/2)}[\psi,\bar\psi,U]=
a^4 \sum_x\,\bar\psi(x) 
\Big{[}\gamma\cdot\widetilde\nabla+
\label{PHYSCHIM}\\&&-i\gamma_5\tau_3W_{\rm{cr}}(r)+m_q\Big{]}\psi(x)
\, .\nonumber\eeqn
The choices $\omega=\pm\pi/2$ are particularly useful because it can 
be proved~\cite{FR} that, despite the fact that the theory is not fully 
O($a$) improved, cancellation of O($a$) effects in quantities of physical 
interest (like energies and operator matrix elements) is either automatic 
or obtainable with no need of any $WA$. The proof of this statement is 
sketched in sect.~\ref{sec:OAI}. O($a$) ambiguities in the knowledge of 
$M_{\rm{cr}}(r)$ do not spoil any of the above results.

Furthermore it is important to observe that at $\omega=\pm\pi/2$ the 
critical WD operator (i.e.\ the operator in square parenthesis in 
eq.~(\ref{PHYSCHIM}) with $m_q=0$) is anti-Hermitian, so its 
spectrum is purely imaginary. This means that the full WD operator cannot
have any vanishing eigenvalue as soon as $m_q\neq 0$. This is also evident 
from the explicit expression of its determinant, ${\mathcal{D}}_{\rm{F,D}}$, 
which takes the remarkably simple form
\beq
{\mathcal{D}}_{\rm{F,D}}={\rm{det}}\Big{[} Q_{\rm cr}^\dagger Q_{\rm cr}
+ m_q^2 \Big{]}\, ,\label{FERDET}\eeq
where $Q_{\rm cr} = Q_{\rm cr}^\dagger$ is defined by
eqs.~(\ref{WDONDD}) and~(\ref{SDRND}).

In Appendix~A we prove that the fermionic action~(\ref{PHYSCHIM}) is 
``stable'' under radiative corrections, in the sense that the unbroken
(possibly spurionic) symmetries of the action prevent radiative corrections 
from generating extra independent operators (of dimension $\leq 4$), not
already present in~(\ref{PHYSCHIM}), which one would need to include for 
renormalizability. Once $M_{\rm cr}(r)$ has been set to the appropriate 
value, the continuum limit is approached as usual by rescaling $g_0^2$ 
and $m_q$ according to the chosen renormalization conditions. 

\subsection{Non-singlet Ward-Takahashi identities}
\label{sec:NSWTI}

In this section we collect the expressions of the renormalized current 
and quark density operators entering the flavour non-singlet Ward-Takahashi 
identities (WTI's) associated with the action~(\ref{PHYSCHIM}). 

Renormalized vector and axial currents can be taken to be 
\beq\begin{array}{ll}
\hat{V}_\mu^1=Z_{A}\,\bar\psi\gamma_\mu\frac{\tau_1}{2}\psi\quad
&\hat{A}_\mu^1=Z_{V}\,\bar\psi\gamma_\mu\gamma_5\frac{\tau_1}{2}\psi\\
\hat{V}_\mu^2=Z_{A}\,\bar\psi\gamma_\mu\frac{\tau_2}{2}\psi\quad
&\hat{A}_\mu^2=Z_{V}\,\bar\psi\gamma_\mu\gamma_5\frac{\tau_2}{2}\psi\\
\hat{V}_\mu^3=Z_V\,\bar\psi \gamma_\mu \frac{\tau_3}{2}\psi\quad 
&\hat{A}_\mu^3=Z_A\,\bar\psi \gamma_\mu \gamma_5\frac{\tau_3}{2}\psi\, ,
\end{array}\label{GENCUR}\eeq
where to make contact with known quantities the (finite) renormalization 
constants, $Z_V$ and $Z_A$, introduced above are those for the local 
vector and axial currents of standard Wilson fermions, respectively. Notice 
the switch between $Z_V$ and $Z_A$ for the currents with flavour $b=1,2$, 
due to the presence of the factor $\gamma_5\tau_3$ in front of the Wilson 
term in eq.~(\ref{PHYSCHIM}) (see also the comment at the end of this section).

With reference to eqs.~(\ref{GENCUR}), the non-singlet WTI's with the 
insertion of the renormalized (multi-local) operator $\hat{O}(y)$ 
($ y\equiv \{y_i,i=1,\ldots,n\}\neq x$) take the expected form ($b=1,2,3$)
\beqn
\hspace{-.5cm}&&\langle{\partial}^\star_\mu\hat{V}^b_\mu(x)
{\hat{O}}(y)\rangle\Big{|}_{(r,m_q)}={\rm{O}}(a) \label{WTITWV}\\ 
\hspace{-.5cm}&&\langle\Big{[}{\partial}^\star_\mu\hat{A}^b_\mu(x)
-2\hat{m}_q\hat{P}^b(x)\Big{]}\hat{O}(y)\rangle
\Big{|}_{(r,m_q)}={\rm{O}}(a)\label{WTITWA}\eeqn
with 
\beqn
&&\hat{m}_q = Z_P^{-1}m_q\, ,\label{RENMA}\eeqn
provided one defines, in terms of bare quantities, the renormalized 
pseudo-scalar operators, $\hat{P}^b$, to be 
\beqn
&&\hat{P}^b=Z_{P}\, \bar\psi\frac{\tau_b}{2}\gamma_5\psi
\qquad b=1,2\label{PTM1}\\
&&\hat{P}^3=Z_{S^0}\Big{[}\bar\psi\frac{\tau_3}{2}\gamma_5\psi +
a^{-3}i\rho_P(am_q) 1\!\!1 \Big{]} \, .\label{PTM2}\eeqn
In eq.~(\ref{PTM2}) the dimensionless real coefficient, $\rho_P(am_q)$, in
front of the power divergent term admits a polynomial expansion in $am_q$. 
Similarly to $Z_V$ and $Z_A$, $Z_P$ and $Z_{S^0}$ are the (logarithmically 
divergent) renormalization constants of the non-singlet pseudo-scalar and 
singlet scalar quark density of the standard Wilson regularization, 
respectively.

For completeness we also give the expression of the renormalized 
singlet scalar quark density operator, $S^0$. By standard symmetry
and dimensionality arguments, one finds 
\beq \hat{S}^0=Z_P\Big[ \bar\psi\psi + 
a^{-2}m_q \rho_{S^0}(am_q) 1\!\!1 \Big] \, ,
\label{RSQD}\eeq
where $\rho_{S^0}$ is a dimensionless real coefficient with an even polynomial 
dependence on $am_q$. This parity property follows from the invariance of the 
action~(\ref{PHYSCHIM}) under the spurionic symmetry 
${\mathcal P}_{\pi/2}^1\times (m_q\rightarrow -m_q)$ (eq.~(\ref{PTILDE_1})).
Formula~(\ref{RSQD}) is rather interesting because it shows that the chiral 
order parameter is only affected by an $m_q/a^2$ power divergence, 
analogously to what happens with Ginsparg-Wilson fermions.
 
The equations of this section can be obtained by specializing to the case 
$\omega_r=\omega=\pi/2$ the corresponding formulae of ref.~\cite{FR}. A 
direct and practical method to deduce them is to proceed in the following way.
One starts by rotating the fermionic fields in the action~(\ref{PHYSCHIM}) 
so as to have the Wilson term in the standard form (i.e. with no factor 
$\gamma_5\tau_3$ in front of it). At this point one can straightforwardly 
adapt to this instance the analysis of the structure of the mixing pattern 
of the chiral rotation of the Wilson term spelled out in detail in 
ref.~\cite{BMMRT} (see also ref.~\cite{T}). In a mass independent 
renormalization scheme renormalization constants can be computed in the 
massless limit ($m_q=0$). For this reason it seemed to us a natural choice 
to keep for them the names they would have in the standard Wilson 
regularization. The final step of this method requires to rotate all the 
quark fields back into the physical basis we started from.

\subsection{O($a$) improvement}
\label{sec:OAI}

It was proved in ref.~\cite{FR} that the invariance of the 
action~(\ref{PHYSCHI}) under
\beqn
&&{\mathcal{R}}_5^{\rm{sp}} =  {\mathcal{R}}_5\times (r\rightarrow -r)\times
(m_q\rightarrow -m_q) \, ,\label{SPUR1}\\
&&\nonumber\\
&&{\mathcal R}_5 \times  {\mathcal D}_d \, ,
\label{SPUR2}
\eeqn
with ${\mathcal R}_5$ and ${\mathcal D}_d$ defined in eqs.~(\ref{R5}) 
and~(\ref{Dd}), respectively, implies the validity of the formula
\beqn
&&\langle O \rangle\Big{|}_{(m_q)}^{(\omega)\,WA} \equiv
\frac{1}{2}\Big{[}\langle O \rangle\Big{|}^{(\omega)}_{(r,m_q)}+
\langle O \rangle\Big{|}^{(\omega)}_{(-r,m_q)}\Big{]}=\nonumber\\
&&=\zeta^{O}_{O}(\omega, r)
\langle O \rangle\Big{|}^{\rm{cont}}_{(m_q)}+{\rm{O}}(a^2)\, ,
\label{IMPR}\eeqn
where $O$ is any gauge invariant m.r.\ (multi-local) operator.

From~(\ref{IMPR}) one can prove a number of physically interesting 
improvement formulae. To see this let us introduce the eigenstates
$|h,n,{\bf k}\>|_{(r,m_q)}^{(\omega)}$ ($h$ and ${\bf k}$ represent the set 
of quantum numbers and three-momenta characterizing the state and $n$ the 
excitation level) of the transfer matrix, $\widehat{T}(\omega,r,m_q)$, with 
eigenvalues $E_{h,n}({\bf k};\omega,r,m_q)$~\footnote{Actually the transfer 
matrix has been constructed only for $|r|=1$. For $|r|<1$ one should make
reference to its square~\cite{OSIYMP}. For simplicity we do not do that 
in the following.}. The eigenvalue equation reads
\beqn
&&\widehat{T}(\omega,r,m_q)
|h,n,{\bf k}\>\Big{|}^{(\omega)}_{(r,m_q)} =\nonumber\\
&&e^{-a E_{h,n}({\bf k};\omega,r,m_q)}
|h,n,{\bf k}\>\Big{|}^{(\omega)}_{(r,m_q)}\, .\label{EVTM}\eeqn
In the notations of ref.~\cite{FR} one gets the formulae
\beqn
\hspace{-.5cm}&&E_{h,n}({\bf k};\omega,r,m_q)+
E_{h,n}({\bf k};\omega,-r,m_q)=\nonumber\\
\hspace{-.5cm}&&\nonumber\\
\hspace{-.5cm}&&=2E_{h,n}^{\rm{cont}}({\bf k};m_q)+{\mbox{O}}(a^2)
\, ,\label{IMPRE}\\\hspace{-.5cm}&&\nonumber\\
\hspace{-.5cm}&&\Big{[}\<h,n,{\bf k}|B|h',n',{\bf k}'\>
\Big{|}^{(\omega)}_{(r,m_q)}+(r\to-r)\Big{]}
%\<h,n,{\bf k}|B|h',n',{\bf k}'\>\Big{|}^{(\omega)}_{(-r,m_q)} = 
\label{IMPRME}\\\hspace{-.5cm}&&=2\zeta^B_B(\omega,r)
\<h,n,{\bf k}|B|h',n',{\bf k}'\>\Big{|}^{\rm{cont}}_{(m_q)}+{\mbox{O}}(a^2)
\, ,\nonumber\eeqn
where $B$ is a gauge invariant m.r.\ local operator. It is important to remark 
that in the whole argument about O($a$) improvement of $WA$'s the twisting 
angle $\omega$ is a totally inert label.

The interesting observation is that at $\omega=\pi/2$ the second term of the
$WA$'s in eqs.~(\ref{IMPRE}) and~(\ref{IMPRME}), i.e.\ the quantities
evaluated with Wilson parameter $-r$, can be rewritten in terms of closely
related quantities but evaluated with the value $r$ of the Wilson parameter. 
As a consequence, O($a$) improved estimates of energies and matrix elements 
can be obtained without having to average results from simulations with 
lattice actions differing by the sign of the Wilson term and critical mass.

Before coming to the proof of this statement, we recall
from ref.~\cite{FR} that at generic values of $\omega$ the product
\beq
{\mathcal P}\times (\omega\to -\omega) \, ,
\label{Pxomin} \eeq 
where ${\mathcal P}$ is the physical parity operator ($x_P=(-{\bf {x}},t)$)
\beqn
{\mathcal {P}}:\left \{\begin{array}{ll}
&\hspace{-.3cm}U_0(x)\rightarrow U_0(x_P)\, ,\\
&\hspace{-.3cm}U_k(x)\rightarrow U_k^{\dagger}(x_P-a\hat{k})\, , k=1,2,3\\
&\hspace{-.3cm}\psi(x)\rightarrow \gamma_0 \psi(x_P)\\
&\hspace{-.3cm}\bar{\psi}(x)\rightarrow\bar{\psi}(x_P)\gamma_0
\end{array}\right . \label{PAROP}\eeqn
is a spurionic symmetry of the tm-LQCD action~(\ref{PHYSCHI}). Since 
$[{\mathcal P}\times(\omega\to -\omega)]^2=1\!\!1$, (multi-local) m.r.\ 
operators can always be taken to have a definite parity, which can be read 
off from the formula   
\beqn 
\langle O^{(p)}(\{x_i\})
\rangle\Big{|}_{(r,m_q)}^{(\omega)}  =\,(-1)^p \,
\langle O^{(p)}(\{x_{iP}\})
\rangle\Big{|}_{(r,m_q)}^{(-\omega)} \, .\nonumber
\eeqn
This relation entails the possibility of defining a notion of parity 
for the eigenstates of the transfer matrix~\cite{FR}. One can, in fact, 
prove the validity of the following equations
\beq
\widehat{\mathcal{P}}\widehat{T}(\omega,r,m_q)\widehat{\mathcal{P}}=
\widehat{T}(-\omega,r,m_q) \, ,\label{PART}\eeq
\beq
\widehat{\mathcal{P}}\,|h,n,{\bf k}\>\Big{|}^{(\omega)}_{(r,m_q)}=
\eta_{h,n}|h,n,-{\bf k}\>\Big{|}^{(-\omega)}_{(r,m_q)}\, ,\label{PARSTATT}\eeq
\beq
E_{h,n}({\bf k};\omega,r,m_q)=E_{h,n}(-{\bf k};-\omega,r,m_q)
\, ,\label{EEM}\eeq
where $\widehat{\mathcal P}$ is the representative of the parity operation 
on the Hilbert space of states of the theory and $\eta_{h,n}$ 
($\eta_{h,n}^2=1$) is what we will call the parity of the state 
$|h,n,{\bf k}\>|_{(r,m_q)}^{(\omega)}$. Indeed $\eta_{h,n}$ can be taken 
to be an $\omega$-independent integer coinciding with the physical parity 
label of the corresponding continuum state. For details we refer the reader 
to Appendix~F of ref.~\cite{FR}. 

It is also immediate to recognize that the action~(\ref{PHYSCHI}) goes  
into itself under the transformation (we recall that $\omega$ is defined 
${\rm mod}~2\pi$)
\beq
(r \to -r) \times (\omega\to \omega \pm \pi) \, .
\label{rxomin} \eeq
If in particular we set, say, $\omega=\pi/2$, the tm-LQCD action takes the 
form~(\ref{PHYSCHIM}) and the invariance~(\ref{rxomin}) becomes 
\beq
(r \to -r) \times (\omega\to -\omega)|_{\omega = \pi/2} \, .
\label{rxomin_spec} \eeq
Either by inspection or by observing that, owing to~(\ref{rxomin_spec}), 
(\ref{Pxomin}) is equivalent to 
\beq
{\mathcal P} \times (r \to -r)  \, ,
\label{Pxomin_spec} \eeq
we conclude that the action~(\ref{PHYSCHIM}) is invariant 
under~(\ref{Pxomin_spec}). This means that at $\omega=\pi/2$ eq.~(\ref{EVTM}) 
and equations from~(\ref{PART}) to~(\ref{EEM}) can be rewritten in the form
(for short we drop the twisting angle label)
\beq
\Big{[}\widehat{T}(r,m_q) -
e^{-a E_{h,n}({\bf k};r,m_q)}\Big{]}|h,n,{\bf k}\>\Big{|}_{(r,m_q)}=0
\label{EVTMP}\eeq
\beq
\widehat{\mathcal{P}}\widehat{T}(r,m_q)\widehat{\mathcal{P}}=
\widehat{T}(-r,m_q) \, ,\label{PARTP}\eeq
\beq
\widehat{\mathcal{P}}\,|h,n,{\bf k}\>\Big{|}_{(r,m_q)}=
\eta_{h,n}|h,n,-{\bf k}\>\Big{|}_{(-r,m_q)}\, ,\label{PARSTATTP}\eeq
\beq
E_{h,n}({\bf k};r,m_q)=E_{h,n}(-{\bf k};-r,m_q)
\, .\label{EEMP}\eeq

Using the last two equations it is now possible to
cast at $\omega=\pi/2$ the $WA$'s~(\ref{IMPRME}) and~(\ref{IMPRE}) 
in a form in which all the relevant lattice data are extracted from
simulations carried out with the action~(\ref{PHYSCHIM}) and a given fixed
value, $r$, of the Wilson parameter. One gets, in fact, the formulae
\beqn
\hspace{-.5cm}&&\<h,n,{\bf k}|B|h',n',{\bf k}'\>\Big{|}_{(r,m_q)}+\nonumber\\
\hspace{-.5cm}&&+\eta_{hnh'n'}^B \<h,n,-{\bf k}|B|h',n',-{\bf k}'\>
\Big{|}_{(r,m_q)} =\nonumber\\\hspace{-.5cm}&&=2\zeta^B_B(r)
\<h,n,{\bf k}|B|h',n',{\bf k}'\>\Big{|}^{\rm{cont}}_{(m_q)}
+{\mbox{O}}(a^2) \, ,
\label{MXE} \eeqn
where $\eta_{hn,h'n'}^B=\eta_{hn}(-1)^{p_B}\eta_{h'n'} = \pm 1$, and
\beqn
&&E_{h,n}({\bf k};r,m_q)+E_{h,n}(-{\bf k};r,m_q)=\nonumber\\
&&\nonumber\\&&=2E_{h,n}^{\rm{cont}}({\bf k};m_q)+{\mbox{O}}(a^2)\, . 
\label{EAVmom} \eeqn
We observe that $\eta_{hn,h'n'}^B$ is the product of the parities of the 
states $|h,n,{\bf k}\>$ and $|h',n',{\bf k}'\>$ (which, if not known, 
can be determined numerically as explained in ref.~\cite{FR}) times the 
parity, $(-1)^{p_B}$, of the local m.r.\ operator $B$.

Notice that in case the lattice matrix element
$\<h,n,{\bf k}|B|h',n',{\bf k}'\>|_{(r,m_q)}$ is invariant under inversion 
of all the external three-momenta (like when all three-momenta vanish), 
the formula~(\ref{MXE}) gets particularly simple. In fact, if 
$\eta_{hn,h'n'}^B=1$, the lattice matrix element turns out to be 
automatically O($a$) improved, while if $\eta_{hn,h'n'}^B=-1$ 
eq.~(\ref{MXE}) implies that the O($a$) improved estimate of 
$\<h,n,{\bf k}|B|h',n',{\bf k}'\>|_{(r,m_q)}$ is zero. This last result 
is in agreement with what one expects in the continuum limit from parity 
invariance. 

An important instance of a quantity which is automatically O($a$) improved 
is $F_\pi$~\cite{FR}, as it can be extracted from two-point correlators 
evaluated at zero external three-momentum.

Eq.~(\ref{EAVmom}) taken at ${\bf{k}}={\bf{0}}$ tells us that masses are
automatically O($a$) improved. 

We end by observing that, although hadronic masses are 
automatically O($a$) improved at $\omega=\pm\pi/2$, it is not possible 
to extract $M_{\rm{cr}}$ from numerical simulations with a discretization 
error which is better than O($a$) (unless the theory is fully O($a$) 
improved \`a la Symanzik). The reason is precisely that at $\omega=\pm\pi/2$ 
the pion mass is automatically O($a$) improved, even if the critical mass is 
known with an O($a$) error (see Appendix~D of ref.~\cite{FR}), thus it 
cannot be sensitive to O($a$) discretization uncertainties in the critical
mass. With an analogous argument a similar conclusion is reached if WTI's 
are used to determine $M_{\rm{cr}}$.

\section{Mass non-degenerate quarks} \label{sec:MNDQ}

The maximally twisted LQCD fermionic action of an $SU_f(2)$ pair of mass 
non-degenerate quark can be conveniently written in the form
\beqn
&&S_{\rm{F,ND}}^{(\pi/2)}[\psi,\bar\psi,U]=
a^4 \sum_x\,\bar\psi(x)\Big{[}\gamma\cdot\widetilde\nabla+
\label{PHYSCHIND}\\&&-i\gamma_5\tau_1 W_{\rm{cr}}(r)
+m_q+\tau_3\epsilon_q\Big{]}\psi(x)
\, ,\nonumber\eeqn
where to keep the mass term real and flavour diagonal we have used the matrix 
$\tau_3$ to split the masses of the members of the doublet. Consequently the 
Wilson term was twisted with the flavour matrix $\tau_1$. Our notations are 
such that $m_q$ and $\epsilon_q$ are both positive.

What we said about the renormalizability of the fermion action in the 
mass degenerate case is valid also here and it is explicitly proved 
in Appendix~A.

\subsection{Non-singlet Ward-Takahashi identities}
\label{sec:NSWTIND}

For mass non-degenerate quarks non-singlet WTI's take 
the continuum-like form $(x\neq y)$
\beqn
\hspace{-.5cm}&&\langle \Big{[}{\partial}^\star_\mu\hat{V}^1_\mu(x)-
2i\hat{\epsilon}_q\hat{S}^2(x)\Big{]}
{\hat{O}}(y)\>\Big{|}_{(r,m_q,\epsilon_q)}={\mbox{O}}(a)\nonumber\\
\hspace{-.5cm}&&\langle \Big{[}{\partial}^\star_\mu\hat{V}^2_\mu(x)
+2i\hat{\epsilon}_q\hat{S}^1(x)\Big{]}{\hat{O}}(y)\>
\Big{|}_{(r,m_q,\epsilon_q)}={\mbox{O}}(a)\nonumber\\
\hspace{-.5cm}&&\langle{\partial}^\star_\mu\hat{V}^3_\mu(x)\,
{\hat{O}}(y)\>\Big{|}_{(r,m_q,\epsilon_q)}=
{\mbox{O}}(a)\nonumber\\
\hspace{-.5cm}&&\nonumber\\
\hspace{-.5cm}&&\langle \Big{[}{\partial}^\star_\mu\hat{A}^1_\mu(x)
-2\hat{m}_q\hat{P}^1(x)\Big{]}{\hat{O}}(y)\>\Big{|}_{(r,m_q,\epsilon_q)}
+{\mbox{O}}(a)\nonumber\\
\hspace{-.5cm}&&\langle \Big{[}{\partial}^\star_\mu\hat{A}^2_\mu(x)
-2\hat{m}_q\hat{P}^2(x)\Big{]}{\hat{O}}(y)\>\Big{|}_{(r,m_q,\epsilon_q)}
={\mbox{O}}(a)\nonumber\\
\hspace{-.5cm}&&\langle \Big{[}{\partial}^\star_\mu\hat{A}^3_\mu(x)
-2\hat{m}_q\hat{P}^3(x)+\nonumber\\
\hspace{-.5cm}&&-\hat{\epsilon}_q\hat{P}^0(x)\Big{]}{\hat{O}}(y)\>
\Big{|}_{(r,m_q,\epsilon_q)}={\mbox{O}}(a)\nonumber\, ,\eeqn
if we make use of the definitions
\beq\begin{array}{ll}
\hat{V}_\mu^1=Z_{V}\,\bar\psi\gamma_\mu\frac{\tau_1}{2}\psi\quad
&\hat{A}_\mu^1=Z_{A}\,\bar\psi\gamma_\mu\gamma_5\frac{\tau_1}{2}\psi\\
\hat{V}_\mu^2=Z_{A}\,\bar\psi\gamma_\mu\frac{\tau_2}{2}\psi\quad
&\hat{A}_\mu^2=Z_{V}\,\bar\psi\gamma_\mu\gamma_5\frac{\tau_2}{2}\psi\\
\hat{V}_\mu^3=Z_A\,\bar\psi \gamma_\mu \frac{\tau_3}{2}\psi\quad 
&\hat{A}_\mu^3=Z_V\,\bar\psi \gamma_\mu \gamma_5\frac{\tau_3}{2}\psi
\end{array}\label{GENCURND}\eeq
\beqn
\hspace{-.5cm}&&\hat{P}^1=Z_{S^0} \Big[ \bar\psi\frac{\tau_1}{2}\gamma_5\psi +
a^{-3}i\rho_P(am_q,a\epsilon_q) 1\!\!1 \Big] \label{PTM1ND}\\
\hspace{-.5cm}&&\hat{P}^b=Z_{P}\,\bar\psi\frac{\tau_b}{2}\gamma_5\psi\, ,
\qquad b=2,3\label{PTMBND}\\
\hspace{-.5cm}&&\hat{P}^0=Z_{S}\,\bar\psi\gamma_5\psi\eeqn
\beqn
\hspace{-.5cm}&&\hat{S}^1=Z_{P^0}\,\bar\psi\frac{\tau_1}{2}\psi
\label{STM1ND}\\
\hspace{-.5cm}&&\hat{S}^2=Z_{S}\,\bar\psi\frac{\tau_2}{2}\psi
\label{STM2ND}\eeqn
\beqn \hspace{-.5cm}&&\hat{m}_q = Z_P^{-1}m_q\qquad \hat{\epsilon}_q=
Z_S^{-1}\epsilon_q\, .\label{RENMAND}\eeqn
As expected, the above formulae turn into eqs.~(\ref{GENCUR}) to~(\ref{PTM2}) 
if we set $\epsilon_q=0$ and perform the cyclic permutation of flavour indices 
$3\to 2\to 1\to 3$. The above WTI's allow us to identify 
\beqn
&&\hat{m}_q^{(+)}=\hat{m}_q+\hat{\epsilon}_q=Z_P^{-1}m_q+Z_S^{-1}\epsilon_q 
\, ,\label{RMP}\\
&&\hat{m}_q^{(-)}=\hat{m}_q-\hat{\epsilon}_q=Z_P^{-1}m_q-Z_S^{-1}\epsilon_q
\label{RMM}\eeqn
as the renormalized masses of the quarks in the doublet.

For completeness we record the formulae
\beqn
&& \hat{S}^0=Z_P\Big[\bar\psi\psi+a^{-2}m_q \rho_{S^0}(am_q,a\epsilon_q)
1\!\!1 \Big] \, \label{STM0ND0}\\
&& \hat{S}^3=Z_S\Big[ \bar\psi \frac{\tau_3}{2} \psi + 
a^{-2} \epsilon_q \rho_{S}(am_q,a\epsilon_q) 1\!\!1 \Big] \, ,
\label{STM0ND3} 
\eeqn
where $\rho_{S^0}(am_q,a\epsilon_q)$ and $\rho_{S}(am_q,a\epsilon_q)$,
as well as $\rho_P(am_q,a\epsilon_q)$, are dimensionless real coefficients
that admit a polynomial expansion both in $am_q$ and $a\epsilon_q$.
The mixing coefficients $\rho_P(am_q)$ and $\rho_{S^0}(am_q)$, appearing in 
eqs.~(\ref{PTM2}) and~(\ref{RSQD}) in the mass degenerate case, coincide 
with the value at $\epsilon_q=0$ of the coefficients introduced in 
eq.~(\ref{PTM1ND}) and~(\ref{STM0ND0}), respectively. 

\subsection{O($a$) improvement}
\label{sec:OAIND}

The method for O($a$) improvement is just as in the mass degenerate case.
The physical explanation of this fact is obvious: all O($a$) discretization
effects come from the Wilson term which is of the same form in the
actions~(\ref{PHYSCHIM}) and~(\ref{PHYSCHIND}) (up to a trivial flavour 
rotation). Formally, it is enough to observe that the 
action~(\ref{PHYSCHIND}) is invariant under the transformations 
${\mathcal R}_5 \times  {\mathcal D}_d$ and
\beqn
\hspace{-.7cm}&&{\mathcal R}_5^{\rm spND} \equiv \label{SPUR1ND} \\
\hspace{-.7cm}&&\equiv{\mathcal R}_5 \times (r \to -r)\times 
(m_q \to -m_q) \times(\epsilon_q \to -\epsilon_q) \, ,\nonumber\eeqn
as well as under the spurionic parity operation~(\ref{Pxomin_spec}).
%\beq {\mathcal P}\times (r\to -r)\label{PARRND} \, .\eeq
Invariance under ${\mathcal R}_5^{\rm spND}$ and 
${\mathcal R}_5\times{\mathcal D}_d$ allows to prove (by arguments 
analogous to those presented in ref.~\cite{FR}) that $WA$'s of energies 
and matrix elements are free from O($a$) cutoff effects. In formulae 
(dropping the label $\omega=\pi/2$)
\beqn
\hspace{-.7cm}&&E_{h,n}({\bf k};r,m_q,\epsilon_q)+
E_{h,n}({\bf k};-r,m_q,\epsilon_q)=
\nonumber\\\hspace{-.7cm}&&\nonumber\\
\hspace{-.7cm}&&=2E_{h,n}^{\rm{cont}}({\bf k};m_q,\epsilon_q)+
{\mbox{O}}(a^2)\, ,\label{IMPREND}\\
\hspace{-.7cm}&&\nonumber\\
\hspace{-.7cm}&&\Big{[}\<h,n,{\bf k}|B|h',n',{\bf k}'\>
\Big{|}_{(r,m_q,\epsilon_q)}+(r\to -r)\Big{]}=\nonumber\\ 
\hspace{-.7cm}&&=2\zeta^B_B(r)
\<h,n,{\bf k}|B|h',n',{\bf k}'\>\Big{|}^{\rm{cont}}_{(m_q,\epsilon_q)}
+{\mbox{O}}(a^2)\, . \label{IMPRMEND}
\eeqn
Using the symmetry of the action~(\ref{PHYSCHIND}) under the 
transformation~(\ref{Pxomin_spec}), one can then prove, along the 
lines of sect.~\ref{sec:OAI}, that averages of energies evaluated
with opposite values of the three-momenta as well as the appropriate linear
combinations of matrix elements of m.r. operators taken between pairs of 
states with opposite external three-momenta are not affected by O($a$) 
discretization errors. In formulae
\beqn
%\hspace{-.7cm}&&E_{h,n}({\bf k};r,m_q,\epsilon_q) =
%E_{h,n}^{\rm{cont}}({\bf k};m_q,\epsilon_q)+ {\mbox{O}}(a^2)
%\, ,\label{EAIND}\\
\hspace{-.7cm}&&E_{h,n}({\bf k};r,m_q,\epsilon_q)+
E_{h,n}(-{\bf k};r,m_q,\epsilon_q)=
\nonumber\\\hspace{-.7cm}&&\nonumber\\
\hspace{-.7cm}&&=2E_{h,n}^{\rm{cont}}({\bf k};m_q,\epsilon_q)+
{\mbox{O}}(a^2)\, ,\label{EAIND}\\
\hspace{-.7cm}&&\nonumber\\
\hspace{-.7cm}&&\<h,n,{\bf k}|B|h',n',{\bf k}'\>
\Big{|}_{(r,m_q,\epsilon_q)}+\nonumber\\
\hspace{-.7cm}&&+\eta_{hnh'n'}^B \<h,n,-{\bf k}|B|h',n',-{\bf k}'\>
\Big{|}_{(r,m_q,\epsilon_q)} = \nonumber\\
\hspace{-.7cm}&&=2\zeta^B_B(r)\<h,n,{\bf k}|B|h',n',{\bf k}'\>
\Big{|}^{\rm{cont}}_{(m_q,\epsilon_q)}+{\mbox{O}}(a^2) \, .\label{MXEND}\eeqn

\subsection{The fermion determinant}
\label{sec:FERDET}

The fermion determinant associated with the action~(\ref{PHYSCHIND}) 
is real and strictly positive, provided 
\beq m_q^2>\epsilon_q^2 \label{INND}\, .\eeq
The proof, though elementary, requires some algebra and is presented 
in detail in Appendix~B.

It is important to realize that~(\ref{INND}) is not a trivial condition, 
because in terms of the renormalized quark masses, $\hat{m}_q^{(\pm)}$, 
it implies the inequality
\beq
\frac{Z_P}{Z_S}>\frac{\hat{m}_q^{(+)}-\hat{m}_q^{(-)}}
{\hat{m}_q^{(+)}+\hat{m}_q^{(-)}}\, .\label{ZPZSIN}\eeq
The latter leads to a rather stringent constraint on the (finite) ratio 
$Z_P/Z_S$, if $\hat{m}_q^{(+)}\gg\hat{m}_q^{(-)}$. One gets, in fact
\beq
\frac{Z_P}{Z_S}>1-2\frac{\hat{m}_q^{(-)}}
{\hat{m}_q^{(+)}}+{\mbox{O}}\Big{[}\Big{(}\frac{\hat{m}_q^{(-)}}
{\hat{m}_q^{(+)}}\Big{)}^2\Big{]}\, .\label{CONZSZP}\eeq
Since numerically one finds~\cite{BHAT,GGRT} for the quenched ratio 
$Z_P/Z_S$ a number somewhat smaller than 1, one might be worried that 
the inequality~(\ref{ZPZSIN}) is not fulfilled. On this issue a 
leverage can be offered by the choice of the value of $r$ which can be any 
real number satisfying $0<|r|\leq 1$. In perturbation theory one finds for 
the standard Wilson action to one-loop~\cite{BMMRT} something like 
$Z_P/Z_S=1-g_0^2 r^2 I(r) +\ldots$, where $I(r)=I(-r)>0$ for $0<|r|\leq 1$ 
and has a finite limit at $r=0$. This suggests that decreasing $|r|$ 
may increase $Z_P/Z_S$. 

\section{Conclusions}\label{sec:CONCL}

In this note we have shown that tm-LQCD at $\omega=\pm\pi/2$ yields a
particularly useful lattice regularization of the Wilson type for the gauge 
theory of mass non-degenerate fermion pairs. Following ref.~\cite{FR}, we
have proved that with this action essentially all physically relevant 
quantities can be evaluated with no O($a$) cutoff effects, while at the 
same time having the fermion determinant real and strictly positive, 
provided $0<\epsilon_q^2<m_q^2$. Monte Carlo simulations are hence safely 
feasible. The difficulties of the standard HMC in the presence of a flavour 
non-diagonal structure of the WD operator, $D_{\rm ND}$ (see 
eq.~(\ref{WDOND})), can be overcome e.g. by stochastically computing its 
determinant using algorithms of the multi-boson~\cite{MB} or 
PHMC~\cite{PHMC} type. Such algorithms should now be based on some 
polynomial approximation of $1/\sqrt{D_{\rm ND}^\dagger D_{\rm ND}}$ 
and naturally allow a simple correction of the employed polynomial 
approximation.

The approach we have discussed can be adapted~\cite{FR2} to the calculation 
of matrix elements of the CP-conserving, $\Delta S=2$ and $\Delta S=1$ 
effective weak Hamiltonian, in a way which we 
expect will solve the problem of ``wrong chirality mixings'' in the 
construction of the corresponding m.r.\ lattice operators. 

\vspace{0.5cm}
{\bf Acknowledgements} - One of us (G.C.R.) would like to thank the 
organizers of LHP2003 for the lively and exciting atmosphere of the 
Conference and its remarkably good organization. This work was supported 
in part by the European Community's Human Potential Programme under contract 
HPRN-CT-2000-00145, Hadrons/Lattice QCD.

\appendix
\section*{Appendix A: Symmetries of the action~(\ref{PHYSCHIND}) and 
absence of counter-terms}
\renewcommand{\thesection}{A}
\label{sec:APPA}

We want to show that the (spurionic) symmetries enjoyed by the fermionic
action~(\ref{PHYSCHIND}) ensure that no extra operators (of dimension 
$\leq 4$) can be generated by radiative corrections, besides those
already present. The discussion that follows applies in particular to the case
$\epsilon_q =0$, implying the ``stability'' under radiative corrections 
of the action~(\ref{PHYSCHIM})~\footnote{The action~(\ref{PHYSCHIND})
at $\epsilon_q =0$ is related to the action~(\ref{PHYSCHIM}) by the harmless
flavour rotation $\psi \to \exp{(-i\tau_2\pi/4)}\psi$, 
$\bar\psi \to \bar\psi \exp{(i\tau_2\pi/4)}$.}.

The list of transformations which leave the action~(\ref{PHYSCHIND}) 
invariant include 

1) lattice gauge transformations, space-time translations and 
hyper-cubic rotations, as well as the $U_V(1)$ vector transformation 
associated to baryon number conservation (none of these will play any 
special role in the discussion)

2) three continuous (one vector, $I^1(\theta_1)$, and two axial, 
$I^{2}(\theta_{2})$, $I^{3}(\theta_{3})$) non-singlet spurionic 
transformations
\beqn
&&{\mathcal{I}}^{1}(\theta_1)\times
(\,m_q +\epsilon_q\tau_3\rightarrow\nonumber\\
&&\rightarrow e^{i\theta_1\tau_1/2}[m_q+\epsilon_q\tau_3] 
e^{-i\theta_1\tau_1/2} \,)\, ,\label{seasym_V1} \\
&&{\mathcal{I}}^{2}(\theta_2) 
\times(\, m_q + \epsilon_q \tau_3 \rightarrow\nonumber\\
&&\rightarrow e^{i\theta_2\gamma_5\tau_3/2} [m_q + \epsilon_q \tau_3] 
e^{i\theta_2\gamma_5\tau_3/2} \, ) \, ,\label{seasym_A2} \\
&&{\mathcal{I}}^{3}(\theta_3) \times 
( \, m_q + \epsilon_q \tau_3 \rightarrow\nonumber\\
&&\rightarrow e^{-i\theta_3\gamma_5\tau_2/2} [m_q + \epsilon_q \tau_3] 
e^{-i\theta_3\gamma_5\tau_2/2} \, )\, ,
\label{seasym_A3} 
\eeqn
where (notice the simplification occuring if $b=1$)
\beqn
{\mathcal{I}}^{b}(\theta):\left \{\begin{array}{ll}
\hspace{-.4cm}&\psi(x)\rightarrow e^{i\gamma_5\tau_1\pi/4}\,
e^{i\theta\tau_b/2}e^{-i\gamma_5\tau_1\pi/4} \psi(x)  \\\nonumber
\hspace{-.4cm}&\bar{\psi}(x)\rightarrow\bar{\psi}(x)e^{-i\gamma_5\tau_1\pi/4}
e^{-i\theta\tau_b/2}\,e^{i\gamma_5\tau_1\pi/4}
\end{array}\nonumber\right  . \nonumber  
\eeqn

3) charge conjugation, ${\mathcal C}$
\beq
{\mathcal C} : \left \{\begin{array}{lll}
\hspace{-.3cm}&U_\mu(x)\rightarrow U_\mu^\star(x) \\
\hspace{-.3cm}&\psi(x)\rightarrow i\gamma_0\gamma_2 \bar\psi^T(x)  \\
\hspace{-.3cm}&\bar{\psi}(x)\rightarrow - i\psi^T(x) \gamma_0\gamma_2
\end{array}\right . \label{CC} 
\eeq

4) the (anti-linear) reflection operation, $\Theta$. It can be 
defined either with respect to the time slice $x_0=0$ (site reflection, 
$\Theta_s$) or $x_0=a/2$ (link reflection, $\Theta_\ell$). Its
action on monomials of fermionic fields is
\beqn
&&\Theta[f(U)\psi(x_1)\ldots\bar\psi(x_n)]=\nonumber\\
&&=f^\star(\Theta [U])\Theta[\bar\psi(x_n)]\ldots\Theta[\psi(x_1)]\, ,
\label{TMON}\eeqn
where $f(U)$ is a functional of link variables and
\beqn
&&\Theta_{s/\ell}[U_k(x)]=U_k^\star(\theta_{s/\ell} x)\, ,\nonumber\\
&&\Theta_{s/\ell}[U_0(x)]=
U_0^T(\theta_{s/\ell} x - a\hat{0})\label{TUUD}\, ,\\
%&&\nonumber\\
&&\Theta_{s/\ell}[\psi(x)]=\bar\psi(\theta_{s/\ell} x)\gamma_0\, ,\nonumber\\
&&\Theta_{s/\ell}[\bar\psi(x)]=
\gamma_0\psi(\theta_{s/\ell} x)\label{TPPB}\, ,\eeqn
with
\beq
%\hspace{-.7cm}
\theta_{\ell}({\bf{x}},t)=({\bf{x}},-t+a)\, ,\quad
\theta_{s}({\bf{x}},t)=({\bf{x}},-t)\label{TETA}
\eeq

5) the pseudo-parity transformations 
\beq
{\mathcal P}_{\pi/2}^1 \times (m_q \rightarrow -m_q) \, , \,
{\mathcal P}_{F}^2 \times (\epsilon_q \rightarrow -\epsilon_q) \, ,\,
{\mathcal P}_{F}^3  \, , \label{seasym_P} \eeq
where ($x_P \equiv (-{\bf x},x_0)$)
\begin{equation}
{\mathcal{P}}_{\pi/2}^1:\left \{\begin{array}{ll}
\hspace{-.3cm}&U_0(x)\rightarrow U_0(x_P)\, ,\\
\hspace{-.3cm}&U_k(x)\rightarrow U_k^{\dagger}(x_P-a\hat{k})\,\, , k=1,2,3\\
\hspace{-.3cm}&\psi(x)\rightarrow i\gamma_5\tau_1 \gamma_0
\psi(x_P)\\
\hspace{-.3cm}&\bar{\psi}(x)\rightarrow i\bar{\psi}(x_P)
\gamma_0  \gamma_5 \tau_1
\end{array}\right . \label{PTILDE_1}
\end{equation}
\begin{equation}
{\mathcal{P}}_{F}^{2,3}:\left \{\begin{array}{ll}
\hspace{-.3cm}&U_0(x)\rightarrow U_0(x_P)\, ,\\ 
\hspace{-.3cm}&U_k(x)\rightarrow U_k^{\dagger}(x_P-a\hat{k})\,\, , k=1,2,3\\
\hspace{-.3cm}&\psi(x)\rightarrow i\tau_{2,3} \gamma_0 \psi(x_P)  \\
\hspace{-.3cm}&\bar{\psi}(x)\rightarrow -i\bar{\psi}(x_P) \gamma_0  \tau_{2,3}
\end{array}\right . \label{PF_23}
\end{equation}

6) the transformation ${\mathcal R}_5 \times {\mathcal D}_d$, where  
\beq
{\mathcal R}_5 :\left \{\begin{array}{ll}
\hspace{-.3cm}&\psi(x)\rightarrow \gamma_5 \psi(x)  \\
\hspace{-.3cm}&\bar{\psi}(x)\rightarrow - \bar{\psi}(x) \gamma_5
\end{array}\right  . \label{R5}\eeq
\beq
{\mathcal{D}}_d : \left \{\begin{array}{lll}    
\hspace{-.3cm}&U_\mu(x)\rightarrow U_\mu^\dagger(-x-a\hat\mu) \\
\hspace{-.3cm}&\psi(x)\rightarrow e^{3i\pi/2} \psi(-x)  \\  
\hspace{-.3cm}&\bar{\psi}(x)\rightarrow e^{3i\pi/2} \bar{\psi}(-x) 
\end{array}\right . \label{Dd} 
\eeq

7) the spurionic transformations~(\ref{Pxomin_spec}), (\ref{SPUR1ND}) and 
${\mathcal{T}}_E\times (r\to -r)$, where ${\mathcal{T}}_E$ is the (euclidean) 
time-axis inversion. The action of ${\mathcal{T}}_E$ on the elementary fields 
of the theory is ($x_T \equiv ({\bf x},-x_0)$)
\beq
{\mathcal {T}}_E :\left \{\begin{array}{ll}
\hspace{-.3cm}&U_0(x)\rightarrow U_0^\dagger(x_T-a\hat{0})\, ,\\
\hspace{-.3cm}&U_k(x)\rightarrow U_k(x_T)\,\, , k=1,2,3\\
\hspace{-.3cm}&\psi(x)\rightarrow \gamma_5\gamma_0 \psi(x_T)  \\
\hspace{-.3cm}&\bar{\psi}(x)\rightarrow \bar{\psi}(x_T) \gamma_0\gamma_5
\end{array}\right  . \label{T}\eeq

\vspace{.3cm}
\noindent In Tables~1 and~2 we list all the independent operators of dimension 
not larger than 4 that cannot appear in the action~(\ref{PHYSCHIND}). 
They are grouped in columns under the name of the corresponding ``killing'' 
symmetry.
\begin{table}
\begin{center}
\caption{We list under the name of the relevant symmetry that forbids them 
all the operators of dimension ${\rm{d}}=4$ and ${\rm{d}}=3$ that cannot 
appear in the density action~(\ref{PHYSCHIND}). By $F\tilde F$ we mean any 
lattice discretization of 
$\epsilon_{\mu\nu\lambda\rho}{\rm Tr}[F_{\mu\nu}F_{\lambda\rho}]$.}
\vspace{.3cm}
\begin{tabular}{|c|c|}
\hline\hline
Dim &${\mathcal{C}}$\\ 
\hline
\vspace{-0.2cm} & \\
${\rm{d}}=4$ & $\bar\psi {\mathcal T} \gamma_5 \gamma\,\widetilde\nabla \psi
\,\, ({\mathcal T} = 1, \tau_1, \tau_3)$ \\ 
\vspace{-0.2cm} & \\
\hline  
\hline
\vspace{-0.2cm} & \\
${\rm{d}}=3$ & $\bar\psi \tau_2\psi$\\
 & $\bar\psi \tau_2\gamma_5\psi$\\
\hline\hline
\end{tabular}
\end{center}\nonumber
%\label{tab:tab1}
%\vspace{.5cm}
\end{table}
The reality properties of the various coefficients in the 
action~(\ref{PHYSCHIND}) are fixed by the invariance under the 
anti-linear reflection operation $\Theta_\ell$ (or $\Theta_s$). 

The conclusion of this discussion is that the form~(\ref{PHYSCHIND}) of the 
action is preserved by radiative corrections and the bare parameters, 
$g_0^2$, $m_q$ and $\epsilon_q$, need only a purely multiplicative 
renormalization. More details on the renormalization of $m_q$ and 
$\epsilon_q$ can be found in the text in sect.~\ref{sec:NSWTIND}. 
%\vspace{-2.0cm}
\begin{table}
\begin{center}
\caption{Continuation of Table~1}
\vspace{.3cm}
\begin{tabular}{|c|c|c|c|}
\hline\hline
Dim &$P_{F}^3$ & $P_{F}^2\times (\epsilon_q\to -\epsilon_q)$\\ 
\hline
\vspace{-0.2cm} & &  \\
${\rm{d}}=4$
& $ \bar\psi {\mathcal T} \gamma\,\widetilde\nabla\psi
\,\, ({\mathcal T} = \tau_1, \tau_2)$ 
&$ \bar\psi \tau_2 \gamma_5 \gamma\,\widetilde\nabla\psi$\\
& $F\tilde F$ &$ \bar\psi \tau_3 \gamma\,\widetilde\nabla\psi $\\
\vspace{-0.2cm} & &  \\
\hline  
\hline
\vspace{-0.2cm} & &  \\
${\rm{d}}=3$  & $ \bar\psi \gamma_5 \psi$& \\
 & $\bar\psi \gamma_5 \tau_3 \psi$ & \\
 & $\bar\psi \tau_1 \psi$ & \\
%%%\vspace{-0.2cm} & & & \\
\hline\hline
\end{tabular}
\end{center}\nonumber
%\label{tab:tab1}
%\vspace{.5cm}
\end{table}

\appendix
\section*{Appendix B: Positivity of the fermion determinant} 
\renewcommand{\thesection}{B}
\label{sec:APPB}

In this Appendix we show that under the condition~(\ref{INND})
the determinant of the WD operator associated with the 
action~(\ref{PHYSCHIND}), namely 
\beq 
D_{\rm{ND}}=\gamma\cdot\widetilde\nabla-
i\gamma_5\tau_1W_{\rm{cr}}(r)+m_q+\tau_3\epsilon_q\, ,
\label{WDOND}\eeq
is real and (strictly) positive. To prove that 
${\mathcal{D}}_{\rm{F,ND}}={\mbox{Det}}[D_{\rm{ND}}]$ is a real number it 
is enough to note the self-adjointness relation
\beq\gamma_5\tau_3D_{\rm{ND}} \gamma_5\tau_3=D_{\rm{ND}}^\dagger\, .
\label{SDR}\eeq
The proof of positivity is somewhat more involved. To proceed it is 
convenient to introduce the auxiliary self-adjoint operator 
\beq Q_{\rm{cr}}=\gamma_5\Big{[}\gamma\cdot\widetilde\nabla+W_{\rm{cr}}(r)
\Big{]}=Q_{\rm{cr}}^\dagger\, ,\label{SDRND}\eeq
in terms of which ${\mathcal{D}}_{\rm{F,ND}}$ can be written in the form
\beq
{\mathcal{D}}_{\rm{F,ND}}={\mbox{Det}}[Q_{\rm{cr}}+i\tau_1m_q+
\gamma_5\tau_3\epsilon_q]\, .\label{DETND}\eeq
In flavour space the operator in the r.h.s.\ of eq.~(\ref{DETND}) is 
represented by the $2\times 2$ matrix
\begin{eqnarray}
\hspace{-.8cm}&&Q_{\rm{cr}}+i\tau_1m_q+\gamma_5\tau_3\epsilon_q=
\gamma_5\,e^{i\pi\gamma_5\tau_1/4}D_{\rm{ND}}\,
e^{i\pi\gamma_5\tau_1/4}\nonumber\\
\hspace{-.8cm}&&\nonumber\\\hspace{-.8cm}&&=\left(\begin{array}{cc}
Q_{\rm{cr}}+\gamma_5\epsilon_q & im_q \\
im_q & Q_{\rm{cr}}-\gamma_5\epsilon_q 
\end{array}\right)\, .
\label{FLM}\end{eqnarray}
For the determinant of this operator one finds
\beqn
&&{\mathcal{D}}_{\rm{F,ND}}={\mbox{Det}}[D_{\rm{ND}}]=\nonumber\\
&&={\mbox{det}}[(Q_{\rm{cr}}+\gamma_5\epsilon_q)
(Q_{\rm{cr}}-\gamma_5\epsilon_q)+m_q^2]=\nonumber\\
&&={\mbox{det}}[Q_{\rm{cr}}^2+m_q^2-\epsilon_q^2+
2\epsilon_q\gamma\cdot\widetilde\nabla]=\nonumber\\
&&={\mbox{det}}[Q_{\rm{cr}}^2+m_q^2-\epsilon_q^2]\cdot
{\mbox{det}}[1+2\epsilon_q B]\, ,\label{DET1}\eeqn
where we have introduced the definition
\beqn
B=X\,\gamma\cdot\widetilde\nabla\,X\, ,\quad
X=(Q_{\rm{cr}}^2+m_q^2-\epsilon_q^2)^{-1/2}\, .\label{DEFX}\eeqn
Since ${\mbox{det}}[Q_{\rm{cr}}^2+m_q^2-\epsilon_q^2]>0$, we only have to 
prove that the second factor in the last equality of~(\ref{DET1}) is a 
positive quantity. To this end we observe that both 
${\mathcal{D}}_{\rm{F,ND}}$ 
(from eq.~(\ref{FLM})) and ${\mbox{det}}[Q_{\rm{cr}}^2+m_q^2-\epsilon_q^2]$ 
(by inspection) are even under $\epsilon_q\rightarrow -\epsilon_q$. We then 
conclude that 
\beq
{\Delta}_{\rm{F,ND}}(\epsilon_q)\equiv{\mbox{det}}[1+2\epsilon_q B]
=\Delta_{\rm{F,ND}}(-\epsilon_q)\, .\label{DET2}\eeq
This property is indeed enough to prove the positivity of $\Delta_{\rm{F,ND}}$
and hence our thesis. Let us consider, in fact, the expansion  
\beq
{\mbox{tr}}[\log(1+2\epsilon_q B)]=\sum_{k=1}^{\infty}\frac{(-1)^{k+1}}{k}\,
(2\epsilon_q)^k\,{\mbox{tr}}[B^k]\, ,\label{DET4}\eeq
valid for sufficiently small values of $\epsilon_q$~\footnote{For instance,
one can take $2|\epsilon_q|$ smaller than the number ${\rm{min}}\,
[m_q,1/||B||_{2\epsilon_q=m_q}]$.}. 
Eq.~(\ref{DET2}) implies that only even powers of $\epsilon_q$ can 
contribute. This fact together with the observation that $B$ is an 
anti-Hermitian operator ($B^\dagger=-B$, because 
$\gamma\cdot\widetilde\nabla$ is anti-Hermitian) allows us to write
\beqn
%\hspace{-.7cm}&&{\mbox{tr}}[\log(1+2\epsilon_q B)]=
\hspace{-.7cm}&&{\mbox{l.h.s. of}}~(\ref{DET4})=
\sum_{n=1}^{\infty}\frac{(-1)^{n+1}}{2n}\,
(2\epsilon_q)^{2n}\,{\mbox{tr}}[(B^\dagger B)^n]=\nonumber\\
\hspace{-.7cm}&&=\frac{1}{2}{\mbox{tr}}[\log(1+4\epsilon_q^2 B^\dagger B)]\, .
\label{DET5}\eeqn
The last equality proves that ${\mbox{tr}}[\log(1+2\epsilon_q B)]$ is a 
non-negative quantity, implying that ${\mbox{det}}[1+2\epsilon_q B]$ is
real and strictly positive.

We conclude with two technical observations: 1) since $B$ is anti-Hermitian, 
its spectrum is purely imaginary, thus $1+2\epsilon_q B$ cannot have any 
vanishing eigenvalue; 2) eq.~(\ref{DET5}), which was proved for sufficiently 
small $\epsilon_q$, can be extended to the actual physical value of the 
mass splitting by analyticity (see the chain of equalities in 
eq.~(\ref{DET1})). 
% main text

\end{document}